\documentclass[aps,pre,reprint]{revtex4-1}

\usepackage{geometry}           % See geometry.pdf to learn the layout options.
\usepackage{graphicx}
\usepackage{amssymb,amsmath}
\usepackage{epstopdf}
\usepackage{array}
\usepackage{multirow}
\usepackage{amsthm}
\usepackage{color}
\usepackage[normalem]{ulem}

\newcommand{\dd}[2]{\frac{d #1}{d #2}}
\newcommand{\pp}[2]{\frac{\partial #1}{\partial #2}}
\newcommand{\grad}{\nabla}
\newcommand{\oneover}[1]{\frac{1}{#1}}
\newcommand{\half}{\frac{1}{2}}
\newcommand{\R}{\mathbb{R}}
\newcommand{\e}[1]{^{(#1)}}
\newcommand{\scr}[1]{\mathcal{#1}}
\DeclareMathOperator{\Tr}{Tr}
\newcommand{\gradt}[1]{\text{grad$_{#1}$}}
\newcommand{\divt}[1]{\text{div$_{#1}$}}
\newcommand{\eps}{\epsilon}

\begin{document}

\title{Stochastic discs that roll}
\author{Miranda Holmes-Cerfon}
\affiliation{Courant Institute of Mathematical Sciences, New York University.}
\date{\today}

\begin{abstract}
We study a model of rolling particles subject to stochastic fluctuations, 
which may be relevant in systems of nano- or micro-scale particles where rolling is an approximation for strong static friction.
We consider the simplest possible non-trivial system: a linear polymer of three of discs constrained to remain in contact, and immersed in an equilibrium heat bath so the internal angle of the polymer changes due to stochastic fluctuations. 
We compare two cases: one where the discs can slide relative to each other, and the other where they are constrained to roll, like gears. 
Starting from the Langevin equations with arbitrary linear velocity constraints, we use formal homogenization theory to derive the overdamped equations that describe the process in configuration space only. 
The resulting dynamics have the formal structure of a Brownian motion on a Riemannian or sub-Riemannian manifold, depending on if the velocity constraints are holonomic or non-holonomic.  
We use this to compute the trimer's equilibrium distribution both with, and without, the rolling constraints. Surprisingly, the two distributions are different. 
We suggest two possible interpretations of this result: either (i) dry friction (or other dissipative, nonequilibrium forces) changes basic thermodynamic quantities like the free energy of a system, a statement that could be tested experimentally, or (ii) as a lesson in modeling rolling or friction more generally as a velocity constraint when stochastic fluctuations are present. In the latter case, we speculate there could be a ``roughness'' entropy whose inclusion as an effective force could compensate the constraint and preserve classical Boltzmann statistics. 
Regardless of the interpretation, our calculation shows the word ``rolling'' must be used with care when stochastic fluctuations are present. 
\end{abstract}

% insert suggested PACS numbers in braces on next line

\maketitle

Particles that live on the nano- or micro-scale %(such as colloids) 
commonly have short-ranged interactions, so their surfaces come close enough that surface frictional effects %, for example due to surface roughness or sticky tethers, 
may be important.  %\cite{Still:2014bd,Mani:2012dia}. %Sircar:2014jc
For example, recent experiments  and simulations have shown that tangential frictional forces between rough, and otherwise stochastic particles, are probably the origin of the shear-thickening behaviour of many materials \cite{Lin:2015gf,Mari:2015fw}. %, yet these particles are also highly stochastic under conditions of low shear. 
Other studies demonstrate that sticky tethers attached to particle surfaces can change their dynamics \cite{Mani:2012dia,Sircar:2014jc}.
Since one promising method of creating colloids with programmable interactions is to coat them with strands of DNA \cite{dreyfus2009,mirkin2011,Rogers:2011et,Rogers:2015bv},  which could impede their relative sliding, this could have major implications for their assembly pathways and hence structures that can be formed by self-assembly.  
On these scales it is extremely difficult to measure the particles' rotational degrees of freedom, so one must resort to indirect methods to determine whether tangential frictional forces are present  \cite{Jenkins:2014js,Still:2014bd}. 
Therefore, it would be highly desirable to find a simpler way to quantify these forces, via macroscopic measurements of spatial positions only. 
%These kinds of particles are heavily studied, partly because of their ability to self-assemble into novel devices and materials (**refs), so it is important to understand the different physical mechanisms at play and their effects on the free energy landscape and assembly pathways of the particles. 

%**or: friction and stochasticity has been studied only in simple, 1d models, although it has been studied in detail in granular... 
While the mascroscopic effect of dry friction has been studied in detail in granular systems \cite{Rivier:2006iv,Taboada:2006ck,Somfai:2007ge,Radjai:2009ki,Liu:2010jx,Estrada:2011ez}, whose components are large and typically athermal, it has rarely been considered for small particles %small enough to be 
subject to thermal fluctuations, except in simple one-dimensional models \cite{Gennes:2005iu,Hayakawa:2005fu,Touchette:2010jf,Menzel:2011iy,Goohpattader:2014kg}. 
A starting point would be to ignore the details of the friction, which are not well understood \cite{Reiter:1994,Vanossi:2013bt}, and consider the limit of infinite friction: stochastic particles that roll relative to each other when they are in contact.
%this can be thought of as a limiting case of infinite stick-slip friction: since friction is a dissipative force that opposes the relative motion of two surfaces in contact, when it is large enough the surfaces cannot move at all relative to each other, except by pivoting about their point of contact. 
Rolling has been studied in non-stochastic systems and is known to produce a wealth of counterintuitive phenomena: a spinning top spontaneously reverses its direction, a golf ball pops out of a hole without hitting the bottom, a dropped quarter spins infinitely quickly in finite time \cite{Gualtieri:2006hj,Tokieda:2013bl,BouRabee:2008bfa}.
Collectively, rolling particles have different phase behaviours than those that slide \cite{Kim:2010cx}.
%\footnote{deleted: Systems with a similar type of constraints (namely non-holonomic) have been studied more generally in robotics, which seeks optimal paths to execute a given task \cite{Bloch:NSbGrpH-,Park:2008fca}. 
%As robots become smaller and smaller -- some functional nanostructures with rolling behaviours are even being made out of individual molecules \cite{Shirai:2006gd} --  the effects of noise or uncertainty will become more and more important. }
Yet despite their intriguing dynamics, %and fascinating mathematical structure, 
rolling has been considered in stochastic settings only for simple %, one-component 
systems such as a rolling ball or sled \cite{Moshchuk:1990hv,Hochgerner:2010fq,Marchegiani:2015fr}, or as a noisy relaxation of the rolling constraint itself \cite{GayBalmaz:2016kh}.

This paper studies a natural model of stochastic, rolling particles, with the aim of determining how rolling could affect quantities that are macroscopically measurable. 
It considers a system whose dynamics can be worked out explicitly: a polymer of three two-dimensional discs that are constrained to roll relative to each other, like gears. Unlike traditional gears, however, the discs can change their relative positions in space. 
We start with the Langevin equations for the stochastic dynamics combined with velocity constraints to model perfect rolling, and from this calculate the equilibrium distribution of the internal angle of the trimer. Surprisingly, the distributions are different depending on if the velocity constraints are included or not. If this is an accurate model of stochastic particles interacting with infinitely strong friction, it suggests that even finite friction could change the free energies of a system of particles. %interacting with finite friction could have different free energies from those that can slide relative to each other. 
Such a result can only hold if the friction force causes the system to deviate from the predictions of classical statistical mechanics, but would be possible to test experimentally via macroscale measurements.

%\comment{Such a result can only hold if the friction force causes the system to deviate from the predictions of classical statistical mechanics. This in turn depends on the details of how friction arises, but would be possible to test experimentally via macroscale measurements. Alternatively, one might view this toy model as a lesson in imposing constraints in a stochastic system, despite the effectiveness of such constraints in non-stochastic, mechanical settings. We speculate that a ``roughness'' entropy could rectify the situation and still allow one to model rolling dynamics ass a constraint. }

 %Therefore particles interacting with finite friction should have different free energies from  those that can slide relative to each other, so measuring the equilibrium occupation probabilities in position space would quantify the importance of friction. 
% To our knowledge this is the first time anyone has considered how rolling, or friction in general, could change the free energy of a colloidal system. 
 %We explain why it is not possible this to predict using traditional tools of statistical mechanics like the Boltzmann distribution. (**maybe not?)
 
An outline is as follows. Section \ref{sec:setup} describes the setup and notation, including the full Langevin equations and specific forms of the constraints for arbitrary collections of discs. Section \ref{sec:overdampshort} describes the overdamped Langevin dynamics. Section \ref{sec:eqb} derives the equilibrium distributions for a trimer of discs both with and without rolling constraints. Section \ref{sec:discussion} discusses the results in a physical context. Section \ref{sec:conclusion} concludes and speculates how this might apply to spheres, whose configuration space is geometrically fundamentally different.

\section{Setup}\label{sec:setup}
%\paragraph{Setup}

\begin{figure}
\center
\includegraphics[width=0.8\linewidth]{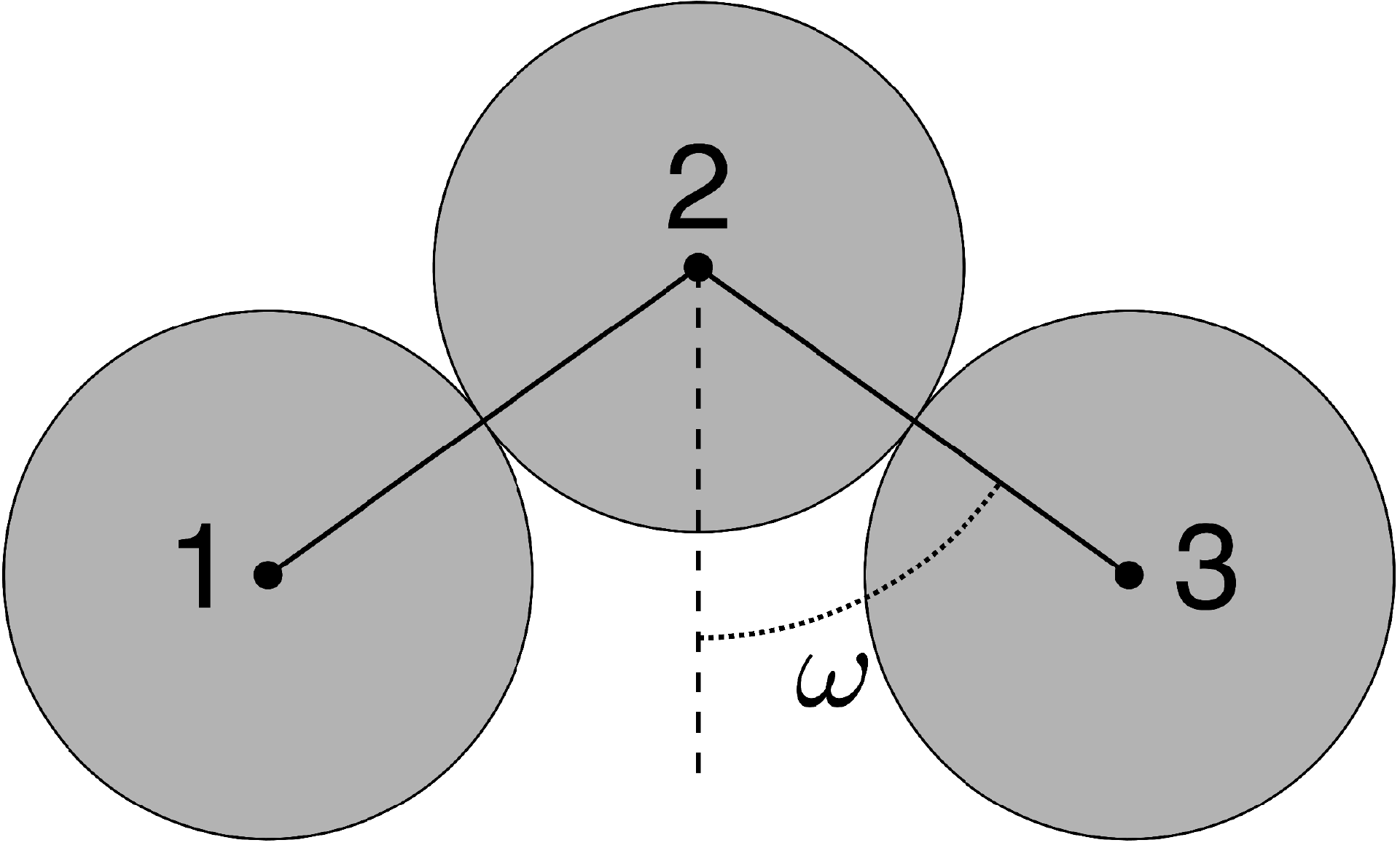}
\caption{A trimer of discs, constrained to preserve the distances between disc pairs 1-2, 2-3, and sometimes constrained to roll when pairs are in contact. This setup illustrates the parameterization in \eqref{eq:param1}.  
}\label{fig:trimer}
\end{figure}

\begin{figure*}
\center
\includegraphics[width=0.9\linewidth]{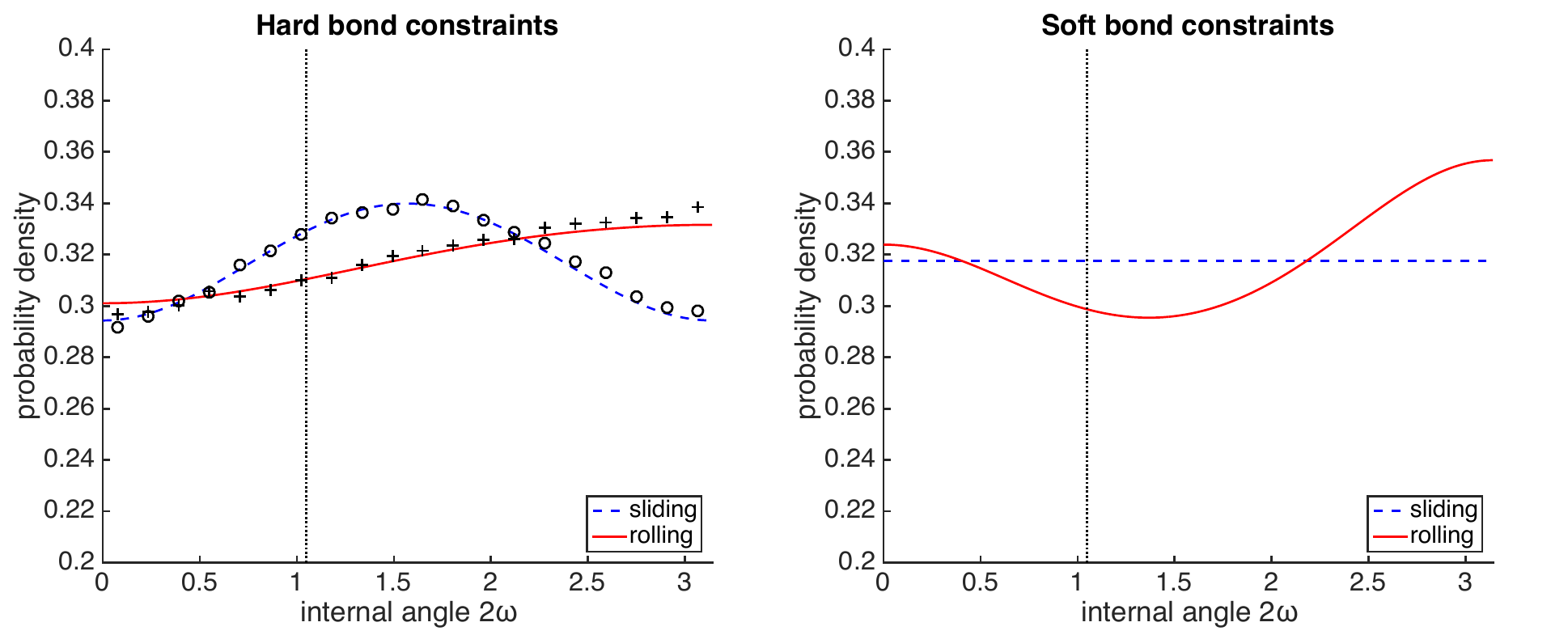}
\caption{
Equilibrium probability densities for a trimer where discs may slide (blue dashed lines) and where they are constrained to roll (red solid lines), as a function of internal angle $2\omega$, in radians. Left plot is for bond constraints imposed exactly, right plot is for bond constraints imposed with a stiff spring. 
Markers are the empirical densities obtained by numerically simulating the Langevin equations \eqref{eq:lang}. 
The vertical dotted line shows where discs 1,3 overlap, so a physical density should be truncated here. 
}\label{fig:probs}
\end{figure*}

We represent the discs as a vector 
$x = (x\e{1},x\e{2},x\e{3},\theta_1,\theta_2,\theta_3)\in \R^{9}$, where each disc has three coordinates representing the center of mass $x\e{i} = (x_i,y_i)$ and the overall internal rotation $\theta_i$ relative to a fixed, external coordinate system. We will call $\{x\e{i}\}_{i=1}^3$ the ``position'' variables because they describe the discs' overall positions in space, and we will call $\{\theta_i\}_{i=1}^3$ the ``spin'' variables because they describe how much each disc has internally rotated, or spun about an axis, like a gear fixed in place. The spin variables are the ones that are usually not accessible by macroscopic measurements. 
All vectors in this paper are column vectors, though we write them inline for readability. 
The discs are identical with unit diameters, and pairs $(1,2)$ and $(2,3)$ are in contact. For each such pair $(i,j)$ there are two possible constraints: one requires the discs to be a fixed distance apart so they are exactly touching,  and another requires the points in contact to move with the same relative velocity. 
These each imply a constraint on the velocities (not momenta), as
\begin{align}
(x\e{i}-x\e{j})\cdot (\dot{x}\e{i}- \dot{x}\e{j}) &= 0  \label{eq:con1}\\
(x\e{i}-x\e{j})^\perp\cdot (\dot{x}\e{i}- \dot{x}\e{j}) &= \frac{1}{2}(\dot{\theta}_i+\dot{\theta}_j)%&=0   
\label{eq:con2}
\end{align}
We write $(u,v)^\perp = (-v,u)$. The second constraint comes from 
noting the velocity on disc $i$ of the point in contact with disc $j$ is 
$\dot{x}\e{i} + \dot{\theta}_i\half(x\e{j}-x\e{i})^\perp$, and 
considering the component of relative velocity that is perpendicular to $x\e{i}-x\e{j}$, since the component parallel to it is accounted for by the first constraint. We call \eqref{eq:con1} the ``bond constraints'' and \eqref{eq:con2} the ``rolling constraints.'' 
In addition, we constrain the center of mass to the origin. %, to make the configuration space compact. 
The complete set of constraints can be written as 
\begin{equation}\label{eq:C}
C(x)\dot{x} = 0,
\end{equation}
where $C(x)\in \R^{m\times n}$ is a matrix whose rows are the coefficients multiplying velocities in  \eqref{eq:con1}, \eqref{eq:con2}.
Here $m=6$ is the number of constraints, and $n=9$ is the number of configuration space variables. 

We suppose the potential energy of the system is a smooth function $U(x)$, and the discs are immersed in a fluid or other medium that provides a white noise forcing to the momentum and a viscous damping that is linear in velocity. We use the Langevin equations to model the dynamics, and
 assume the friction tensor $\Gamma(x)$  and forcing tensor $\sigma(x)$ satisfy a fluctuation-dissipation relation $\sigma\sigma^T = 2\beta^{-1}\Gamma$, where $\beta = (k_BT)^{-1}$ is the inverse of temperature $T$ times the Boltzmann constant. This ensures that the invariant measure for the unconstrained system is the Boltzmann distribution: $e^{-\beta U(x)}e^{-\frac{\beta}{2}\dot{x}^TM\dot{x}}$. Here $M$ is the mass matrix, which is diagonal with entries equal to either the mass or moment of inertia of a disc.
 
It should be noted that the Langevin equations do not correctly describe the velocity correlations of particles immersed in a fluid, even for large particles, because the correlation times for the particles' velocities and for the fluid's momentum fluctuations are of the same magnitude \cite{Hinch:1975ba,Roux:1992ic}. 
 However, they do capture the correct static thermodynamic behavior (at least without constraints), and they do lead to the correct overdamped equations. Since our goal is to obtain an overdamped equation which involves restrictions on velocities, and not to correctly describe the velocity correlations induced by hydrodynamic interactions, we proceed with the Langevin equations as a starting point from which we can impose velocity constraints via mechanical principles. 

To account for constraints, we apply d'Alembert's principle, or the principle of virtual work. This requires that the constraints do no work in a ``virtual'' move, namely one which holds all variables fixed and takes a step in a tangent direction consistent with the constraints. The constraints must therefore be imposed by forces perpendicular to the allowable tangent directions, which can be done using Lagrange multipliers \cite{Landau:1976tn}. This is the only principle available for arbitrary linear velocity constraints, since variational principles are only valid when the constraints are known to be holonomic \cite{Flannery:2005bi}. The constrained Langevin equations are
\begin{equation}\label{eq:lang}
M\ddot{x} + \Gamma \dot{x} = - \grad U(x) +  \sigma \eta(t)- C^T\lambda 
\end{equation}
combined with the constraints \eqref{eq:C}. In the above, $\eta(t)$ is a $n$-dimensional white noise, and 
 $\lambda\in\R^{n}$ are the Lagrange multipliers that ensure the constraints are satisfied. 
 The product $\sigma(x)\eta$ can be interpreted in either the It\^o or the Stratonovich sense, since $\sigma(x)$ does not depend on $\dot{x}$ so it is of bounded variation. 
 The mass can be removed from the equations by changing to mass-scaled variables (see appendix \ref{sec:mass}, or \cite{Lelievre:2010uu}), so hereafter we set $M=I$. This is not a non-dimensionalization, but simply a convenient change of variables, which can be inverted to put the mass back in at any step in the subsequent analysis when desired.

 \section{Overdamped dynamics}\label{sec:overdampshort}
 
The particles we aim to model have very short correlation times for momentum, so they are effectively modelled %on experimentally measurable timescales 
by the overdamped Langevin equations, which describe the dynamics of the system in configuration space only. We derive these overdamped equations by considering the limit of large viscous friction and long timescales.  In this section we sketch the results; the detailed calculations are shown in appendices  \ref{sec:lagrange}, \ref{sec:overdamp}. 
 
First, we write \eqref{eq:lang} explicitly. 
The Lagrange multipliers can be computed analytically by taking the time derivative of \eqref{eq:C} and substituting for $\ddot{x}$ from \eqref{eq:lang} (see appendix \ref{sec:lagrange}.) The resulting equations are
\begin{multline}\label{eq:lang2}
\ddot{x} + P\Gamma\dot{x} =\\ -P\grad U(x) + P\sigma \eta(t)+ C^TG^{-1}\grad C(\dot{x},\dot{x}).
\end{multline}
The matrix $P(x)$ is an orthogonal projection onto the complement to the row space of $C(x)$, and $G(x)$ is the Gram matrix of the constraints:
\begin{equation}\label{eq:P}
P = I - C^TG^{-1}C,  \qquad G = CC^T.
\end{equation}
Here and hereafter $I$ is an identity matrix with dimensions correct for the context. 
The final term in \eqref{eq:lang2} is a vector with components $(\grad C(x)(\dot{x},\dot{x}))_i = \sum_{i,j}\pp{c_i}{x_j}\dot{x}_i\dot{x}_j$, %\dot{x}^T(\grad c_{i1},\ldots,\grad c_{im})\dot{x}$, 
and represents the extra acceleration due to the curvature of the constraints. 
This shows the constrained dynamics are given by projecting the original momentum equation onto the subspace of allowed, unconstrained directions, plus a curvature-driven acceleration term \cite{Ciccotti:2007fv}. Again it doesn't matter if the projected force $P(x)\sigma(x) \eta(t)$ is interpreted in the It\^o or Stratonovich sense because $x$ is of bounded variation.

Next, we consider the overdamped limit by letting $\Gamma \to \Gamma/\epsilon$ and $t\to t/\epsilon$, and performing formal homogenization on the generator of \eqref{eq:lang2} \cite{Pavliotis:2008wg}. This is a standard technique to obtain the overdamped Langevin equations asymptotically; the novelty here is the arbitrary linear velocity constraints.
The result is weakly equivalent to the stochastic process $x(t)$ that solves the It\^o equation (see appendix \ref{sec:overdamp} for details) 
\begin{equation}\label{eq:xbar}
\dot{x} = -\Gamma_P^\dagger\grad U - \beta^{-1}\sum_{ij}P_{ij}\partial_j(\Gamma_P^\dagger)_{ki} + \bar{\sigma}_P\eta.
\end{equation}
Here $\Gamma_P= P\Gamma P$, and $\Gamma_P^\dagger$ is its Moore-Penrose pseudoinverse \footnote{Given an $m\times n$ matrix $A$, the Moore-Penrose pseudoinverse  $A^\dagger$ is the unique matrix which satisfies (i) $AA^\dagger A = A$, (ii) $A^\dagger AA^\dagger = A^\dagger$, (iii) $(AA^\dagger)^T = AA^\dagger$, and (iv) $(A^\dagger A)^T=A^\dagger A$.}  \cite[see e.g.][]{Strang:va}. %It can be computed from the singular value decomposition $A=U\Sigma V^T$ as $A^\dagger=V\sigma^\dagger U^T$.
The matrix $\bar{\sigma}_P$ is any matrix such that $\bar{\sigma}_P\bar{\sigma}_P^T = 2\beta^{-1}\Gamma_P^\dagger$.

To highlight the fundamental ideas we will analyze \eqref{eq:xbar} in the simplest possible setup: constant friction and no long-range potential energy, so $\Gamma=I$ and $U(x)=0$.
In this case $\Gamma_P^\dagger = P$, so after a change of time scale $t\to t/(2\beta)$  \eqref{eq:xbar} becomes
\begin{equation}\label{eq:dx}
\dot{x}= %-\half\sum_{ij}P_{ij}\partial_jP_{ki}dt + Pd\eta = 
P(x)\circ \eta ,
\end{equation}
where $\circ$ denotes the Stratonovich product. 
The process $x(t)$ looks locally like a Brownian motion that can only move in a subspace of its ambient space. 

This process is well-understood mathematically when the velocity constraints are holonomic, meaning they imply an equal number of constraints in configuration space, i.e. on the variables contained in $x$. In this case the process is constrained to remain on a manifold $\mathcal M$ in configuration space, whose dimension equals the rank of $P(x)$ (which is an orthogonal projection matrix onto the tangent space to $\mathcal M$ at each point $x$.) The process is actually a Brownian motion on $\mathcal M$,  which is by definition a process whose generator is the Laplace-Beltrami operator on $\mathcal M$ \cite{Ikeda:1981,Hsu:1988fl}.
Briefly, to see why, note that the generator of \eqref{eq:dx} is 
\begin{equation}\label{eq:Lu}
\mathcal L u = \half \Tr(P\grad(P\grad u)) %= \half P_{jk}\partial_jP_{ik}\partial_i u + \half P_{ij}\partial_{ij}u
\end{equation}
where $\Tr(AB) = A{:}B= \sum_{i,j}A_{ij}B_{ij}$ for matrices $A,B $, and $\grad$ applied to a vector expands each element into a row so that $(\grad v)_{ij} = \partial_j v_i$. 
Here $u$ may be thought of as a function on $\mathcal M$, even though the gradient operator in $\mathcal L$ acts on all directions in the ambient space. 
Then, $P\grad u$ can be shown to equal $\gradt{} u$ where $\gradt{}$ is the gradient operator on $\mathcal M$, and $\Tr (P\grad v)$ can be shown to equal $\divt{} v$, where $\divt{}$ is the divergence operator on $\mathcal M$ \cite[e.g.][]{Hartmann:2005dh,Ciccotti:2007fv,Lelievre:2010uu}. Therefore for holonomic constraints $\mathcal L u = \half\divt{}\gradt{} u$, which is the Laplace-Beltrami operator on $\mathcal M$. 

For non-holonomic constraints there is no similarly canonical interpretation of $\mathcal L$ of the form \eqref{eq:Lu} at the present time, as we discuss briefly in the conclusion. 

For discs, the bond constraints \eqref{eq:con1} are holonomic since they imply the distances between pairs in contact are conserved, which is a constraint in configuration space. The constraints on the center of mass are also holonomic. The rolling constraints \eqref{eq:con2}  are not immediately seen to be holonomic, since they cannot be integrated in time directly. Although one can show they are indeed holonomic as we discuss briefly in section \ref{sec:hardsoft} and appendix \ref{sec:geometry}, we will proceed without this knowledge, to show that one can still work with \eqref{eq:dx} without knowing the geometric structure of the constraints. 

%In the simplest possible setup with constant friction and no long-range potential energy, and after a change of time scale $t\to t/(2\beta)$,
%the result is weakly equivalent to the process
%\begin{equation}\label{eq:dx}
%\dd{X_t}{t}= %-\half\sum_{ij}P_{ij}\partial_jP_{ki}dt + Pd\eta = 
%P(X_t)\circ \eta .
%\end{equation}
%where $\circ$ denotes the Stranonovich product. (The general result is shown in the appendix.) 

\section{Equilibrium distribution}\label{sec:eqb}
%\paragraph{Equilibrium probability}

\subsection{Result}
%\paragraph{Result}
Next we ask what is the equilibrium distribution for a trimer in position space both with, and without, the rolling constraints.
 We will show these have densities proportional to, respectively,%\footnote{Note that in an arc-length parameterization of internal angle, i.e. in a variable $s$ such that $\pp{x}{s} \parallel \pp{x}{\omega}$, $|\pp{x}{s}|=1$, the stationary distributions would be divided by a factor of $\sqrt{1+2\cos^2\omega}$ (this is the length of $\pp{x}{\omega}$.) Therefore that for sliding would be $\pi_{\text{slide}}(s) = \sqrt{1+2\sin^2\omega(s)} = I(s)$, where $I$ is the two-dimensional moment of inertia of the cluster. }
\begin{align}\label{eq:pitrimer}
\pi_{\text{slide}}(\omega) &\propto \sqrt{1+2\sin^2\omega}\sqrt{1+2\cos^2\omega}, \nonumber \\
\pi_{\text{roll}}(\omega) &\propto  \sqrt{5+2\sin^2\omega}\sqrt{13+2\cos^2\omega},
\end{align}
where $2\omega$ is the internal angle of the trimer. 
The domain is $\omega \in [\pi/6,5\pi/6]$ if the spheres cannot interpenetrate, and $\omega\in[0,\pi]$ if they can (as is allowed in simulations.) 
These calculations are performed for the simplest setup described by \eqref{eq:dx}, but we expect them to be valid in more general settings (with a suitable modification to account for the potential energy $U(x)$.) 

Figure \ref{fig:probs} plots the two distributions. The rolling constraints favour more open configurations than purely bond constraints. This figure also plots the empirical histograms obtained by numerically simulating the Langevin equations \eqref{eq:lang} directly (see appendix \ref{sec:numerics} for methods); the agreement verifies our calculations. The small discrepancies are thought to arise partly from statistical fluctuations, and, in the case of rolling constraints, because the numerical method does not conserve the additional  implied constraints in configuration space (see appendix \ref{sec:geometry}.)

%The sliding trimer has an angle distribution which is peaked at intermediate angles. The rolling trimer has a smoother distribution: it has lowest probability for small angles, where discs sit on top of each other, and highest probability for large angles, where it is nearly collinear. At large angles, motion along a trimer's internal degrees of freedom is most effectively converted to rolling. 

The distributions above are for ``hard'' constraints,  i.e. the constraints are satisfied exactly. In a physical system constraints are often an approximation for a concentration of probability near a lower-dimensional manifold, but the system can wiggle around near this:  the constraints are ``soft''. 
This happens, for example, when constraints of the form $q_i(x)=z$ (where $i$ indexes the constraints) are imposed by a stiff potential energy, such as $U(x) = \eps^{-1}|q_i(x)-z|^2$ with $\eps \ll 1$. 
This wiggle room changes the equilibrium density, 
and in the limit of infinite stiffness it is not the same as imposing hard constraints; this is the well-known ``paradox'' of hard versus soft constraints in statistical mechanics that has been discussed many times in the literature \cite[e.g.][]{Fixman:1974dd,Hinch:1994ca}. The distributions for infinitely stiff soft constraints can be obtained from those for hard ones and we will show they are 
\begin{align}\label{eq:pitrimervibr}
\pi_{\text{slide,vibr}}(\omega) &\propto 1,\nonumber \\
\pi_{\text{roll,vibr}}(\omega) &\propto  \sqrt{\frac{5+2\sin^2\omega}{1+2\sin^2\omega}}\sqrt{\frac{13+2\cos^2\omega}{1+2\cos^2\omega}}. 
\end{align}
These are the distributions one would typically compare to experimentally; for example similarly obtained distributions accurately predict the equilibrium probabilities of colloidal clusters \cite{Perry:2015ku}. 
The distribution when discs can slide is constant (see Figure \ref{fig:probs}), as one would expect since each outer disc should be uniformly distributed on the surface of the central disc.

\subsection{Derivation}\label{sec:derivation}

In this section we show \eqref{eq:pitrimer},\eqref{eq:pitrimervibr} explicitly. This section is technical and not essential to understanding the subsequent discussion.  

Our strategy will be to parameterize the position degrees of freedom of the cluster explicitly to remove the bond and center of mass constraints, write the equations in these variables, and finally solve the Fokker-Planck equation by direct calculation. This is a brute-force approach yet it still gives insight into the geometry and mechanics of the constraints, by explicitly identifying the linear subspaces involved in setting the dynamics. 

\subsubsection{Hard constraints}
 
 Let $\omega \in [0,\pi]$ (or $[\pi/6,5\pi/6]$) be half the internal angle, measured underneath the line 1-2-3 when disc 2 has been rotated to lie on the $y$-axis, %, though it will not end up mattering since the dynamics are symmetric under the transformation $\omega \to \pi-\omega$.
and let $\phi\in[0,2\pi]$ be the overall rotation of the cluster. 
See Figure \ref{fig:trimer} for an illustration. 
Let the position variables be 
\begin{equation}
\bar{x}(\phi,\omega) = R(\phi)\bar{x}_0, 
\end{equation}
where 
\begin{multline}\label{eq:param1}
\bar{x}_0 = \\
(-\sin \omega, -\oneover{3}\cos\omega, 0, \frac{2}{3}\cos\omega,\sin\omega,-\oneover{3}\cos\omega)
\end{multline}
and $R(\phi)$ is a $6\times 6$ block diagonal matrix, whose blocks are $2\times 2$ matrices that rotate each point $x\e{i}$ by an angle $\phi$ about the origin \footnote{Specifically, the blocks are
%\begin{equation*}
$R_1(\phi) = \begin{pmatrix}
\cos\phi & -\sin\phi \\ \sin\phi & \cos\phi
\end{pmatrix}$.}.
%\qquad \text{so that }\quad
%\dd{}{\phi}R_1(\phi) = \begin{pmatrix}
%-\sin\phi & -\cos\phi \\ \cos\phi & -\sin\phi
%\end{pmatrix}.
%\end{equation*}}.
The full cluster is parameterized by 
\begin{equation}\label{eq:param2}
x = (\bar{x},\theta) = (R(\phi)\bar{x}_0, \theta_1,\theta_2,\theta_3).
\end{equation}
This preserves the bond and center of mass constraints so they can be removed from the rows of $C(x)$ which form the projection $P(x)\in \R^{9\times 9}$. %If we write $\mathcal M$ for the set of configurations that are accessible while maintaining disc pairs $\{1,2\}$, $\{2,3\}$ in contact and the center of mass at the origin, then $y = (\omega,\phi, \theta_1,\theta_2,\theta_3)$ is a chart for the manifold $\mathcal M$. 

We now perform this change of variables in \eqref{eq:dx}, to write the dynamics in terms of the new variables $y = (\omega,\phi, \theta_1,\theta_2,\theta_3)$. 
Let $s=(\omega,\phi)$ be the position variables and let $\theta = (\theta_1,\theta_2,\theta_3)$ be the spin variables. 
Let us define the following matrices: 
\begin{equation}\label{eq:ST}
\begin{array}{ll}
S = \grad_s x = (\pp{x}{\omega},\pp{x}{\phi}) & \in \R^{9\times 2}\\
T = \grad_\theta x = (\pp{x}{\theta_1},\pp{x}{\theta_2},\pp{x}{\theta_3})& \in \R^{9\times 3}\\
Y = \grad_y x = (S \quad T)& \in\R^{9\times 5}\\
Q = S^TS = \begin{pmatrix} K^2 & 0 \\ 0 & L^2 \end{pmatrix} & \in\R^{2\times2}
\end{array}
\end{equation}
Here $0$ is the matrix of zeros with appropriate dimensions, and the diagonal elements of $Q$ are
\begin{align}
K^2(\omega) &= \Big| \pp{\bar{x}}{\omega} \Big|^2 = \frac{2}{3} + \frac{4}{3}\cos^2\omega, \label{eq:K}\\
L^2(\omega) &= \Big| \pp{\bar{x}}{\phi}\Big|^2 = \frac{2}{3} + \frac{4}{3}\sin^2\omega. \label{eq:L}
\end{align}
Note that $L^2(\omega)$ is the two-dimensional moment of inertia of the cluster. 

%The geometric interpretation of these matrices is as follows: the matrix $S\in \R^{9\times 2}$ contains the tangent vectors to the position degrees of freedom on $\mathcal M$, $T\in \R^{9\times 3}$ contains the tangent vectors to the rotation degrees of freedom on $\mathcal M$, and $Y\in\R^{9\times 5}$ is the entire set of tangent vectors on $\mathcal M$. 
%The matrix $Q\in\R^{2\times2}$ can be interpreted as the metric tensor on the manifold $\mathcal N$ of accessible position configurations, i.e. the manifold which ignores the rotation degrees of freedom, or sets them to be fixed constants \cite{Kuhnel:2002vh}.

We can use the regular chain rule of calculus on \eqref{eq:dx}, since this is in Stratonovich form. This gives  
\begin{equation}
\dd{x}{t} = \sum_{i=1}^5 \pp{x}{y_i}\dd{y_i}{t}= Y\dd{y}{t} = P(x)\circ \eta.
\end{equation}
Multiplying by $(Y^TY)^{-1}Y^T$ gives an equation for $\dot{y}$.
Note that $Y^T Y = \begin{pmatrix} Q & 0 \\ 0 & I \end{pmatrix}$, since $S^TT =0$, $T^TS = 0$, and $T^T T = I$. %, so $Y^TY$ is invertible.
Separating the equations for the position and spin variables separately gives 
\begin{align}
\dot{s} &= Q^{-1}(PS)^T\circ \eta  \label{eq:ds}\\
\dot{\theta} &= (PT)^T\circ \eta\label{eq:dtheta}
\end{align} 
Here $\eta\in\R^{9}$ is the same white noise for each. 

Notice that equation  \eqref{eq:ds} for the position variables does not depend on $\theta_i$, because $P,S,Q$ are independent of $\theta_i$. (The spin variables, however, do depend on the positions.) 
Therefore we can analyze it independently, to compute the equilibrium distribution in these variables only.

First consider the equilibrium density for \eqref{eq:ds} \emph{without} the rolling constraints, so that 
 $P=I$. 
 One strategy would be to compute the matrix elements in \eqref{eq:ds} directly and solve the stationary Fokker-Planck equation, as we will do when rolling constraints are included. However, it is simpler to proceed geometrically, and recognize that, based on \eqref{eq:dx} and the subsequent discussion, \eqref{eq:ds} is a parameterized version of a Brownian motion on a manifold. (Note that $S$ has zeros in the entries corresponding to the spin variables so these components of $\eta$ do not contribute.) This manifold (call it $\mathcal N$) is the set of accessible configurations in position space when internal rotations are ignored. It can be embedded in the full configuration space by setting the spin variables to fixed constants, for example as $\mathcal N = \{x\!\!: \theta_1\!\!=\!\!\theta_2\!\!=\!\!\theta_3\!\!=\!\!0\}$. This embedding respects the inner product inherited from the ambient space, so the columns of $S$ form a basis for tangent vectors to $\mathcal N$ and $Q$ is the metric tensor on $\mathcal N$ in the variables $(\omega,\phi)$ \cite{Kuhnel:2002vh}. 
The equilibrium density is the surface measure on $\mathcal N$, which in these variables is
$\pi_{\text{slide}}(\omega,\phi) = |Q|^{1/2}$. 
Result \eqref{eq:pitrimer} follows from \eqref{eq:ST},\eqref{eq:K},\eqref{eq:L}.

Next consider the invariant measure \emph{with} the rolling constraints. 
The stationary probability density $\pi_{\text{roll}}(\omega,\phi)$ for \eqref{eq:ds} solves the stationary Fokker-Planck equation 
%\begin{equation}\label{eq:FPB}
%\partial_i \left( B_{ik}\partial_j(B_{jk} p)\right) = 0
%\end{equation}
%with $B = G^{-1}(PS)^T$, 
\begin{equation}\label{eq:FPij}
\sum_{i,j=1}^2\partial_i (c_{ij} \pi_{\text{roll}}+ d_{ij}\partial_j \pi_{\text{roll}}) = 0, 
\end{equation}
where $c_{ij} = b_i\cdot \partial_j b_j$, $d_{ij} = b_i\cdot b_j$,
 and $b_i$ is the $i$th row of the matrix $B= Q^{-1}(PS)^T\in \R^{2\times 9}$. Here $\partial_1=\partial_\omega$, $\partial_2 = \partial_\phi$. 
 The boundary condition is the one which conserves probability: a no-flux boundary condition in $\omega$ which requires $\sum_j(c_{1j} \pi_{\text{roll}}+ d_{1j}\partial_j \pi_{\text{roll}})=0$ at $\omega=\pi/6,5\pi/6$ (or $\omega=0,\pi$), and a periodic boundary condition in $\phi$. 
To determine $b_i$ we first compute an orthonormal basis of $P$, as: 
\begin{align}
t_r &= (0,\ldots,0,1,-1,1)/\sqrt{3}  \nonumber\\
t_\omega &= (\pp{\bar{x}}{\omega},-2,0,2) / \sqrt{K^2 + 8}  \label{eq:horizontal}\\
t_\phi &= (\pp{\bar{x}}{\phi}, \frac{2}{3},\frac{4}{3},\frac{2}{3}) / \sqrt{L^2+ \frac{8}{3}}. \nonumber
\end{align}  

These are obtained as follows: $t_r$ is the motion obtained by fixing the positions of the discs and only letting them spin; we call this ``pure spinning.'' 
For $t_\omega$ we prescribe the first six components to be $\pp{\bar{x}}{\omega}$, and solve the two linear equations \eqref{eq:con2} for $\dot{\theta}$. There is a one-parameter family of solutions $\dot{\theta} = (-2,0,2) + \dot{\theta}_2(-1,1,-1)$. We choose the one which minimizes $|\dot{\theta}|^2$, or equivalently which is perpendicular to $t_r$.
For $t_\phi$ we similarly fix the first six components to be $\pp{\bar{x}}{\phi}$ and solve for $\dot{\theta}$. The solutions are $\dot{\theta} = (0,2,0) + \dot{\theta}_1(1,-1,1)$, and we choose the one with minimum $L_2$-norm. 
Each set of solutions for $\dot{\theta}$ has physical meaning since they tell us how the discs must spin, like gears, to produce a desired motion of the cluster in position space. They are each equal to a fixed combination of spins ($\dot{\theta}=(-2,0,2)$ to change the internal angle, and $\dot{\theta}=(0,2,0)$ to rotate the cluster overall), plus an arbitrary multiple of the pure spinning motion $t_r$. 

We project each column of $S$ using \eqref{eq:horizontal} to find 
$P\pp{x}{\omega} = \frac{K^2}{\sqrt{K^2+8}}t_\omega$, $P\pp{x}{\phi} = \frac{L^2}{\sqrt{L^2+8}}t_\phi$, so 
%\begin{align}
%b_1 &=t_\omega^T/\sqrt{K^2+8},\nonumber\\ 
%b_2 &= t_\phi^T/\sqrt{L^2+8/3}.
%\end{align}
\begin{align}
b_1 =\frac{t_\omega^T}{\sqrt{K^2+8}},\quad
b_2 = \frac{t_\phi^T}{\sqrt{L^2+8/3}}.
\end{align}
%\footnote{
%\begin{equation}
%PS = \begin{pmatrix} 
%\frac{K^2}{\sqrt{K^2+8}} & 0 \\
%0 & \frac{I^2}{\sqrt{I^2+8}}
%\end{pmatrix}
%\qquad 
%\Longrightarrow
%\quad 
%b_1 = \frac{t_\omega^T}{\sqrt{K^2+8}}, \qquad 
%b_2 = \frac{t_\phi^T}{\sqrt{I^2+8/3}}.
%\end{equation}
%}
From this, computing the $c_{ij}$, $d_{ij}$ in \eqref{eq:FPij} is a matter of algebra. 
We eventually write \eqref{eq:FPij} as 
\begin{multline}\label{eq:FPeta}
\partial_1\!\left( ( \alpha_1\alpha_1' - LL'\alpha_1^2\alpha_2^2)\pi_{\text{roll}} +  \alpha_1^2\partial_1 \pi_{\text{roll}}\right) \\+ \partial_2(\alpha_2^2\partial_2\pi_{\text{roll}}) = 0,
\end{multline}
where $'$ denotes a derivative with respect to $\omega$, and 
\begin{align}
\alpha_1(\omega) = |b_1| &= (K^2(\omega)+8)^{-1/2}\nonumber\\
\alpha_2(\omega) = |b_2| &= (L^2(\omega)+8/3)^{-1/2}. \label{eq:alph}
\end{align} 
A solution that is independent of $\phi$ is $\pi_{\text{roll}}(\omega)\propto(\alpha_1\alpha_2)^{-1}=(K^2(\omega)+8)^{1/2}(L^2(\omega)+8/3)^{1/2}$. One can check this satisfies the boundary conditions, so \eqref{eq:pitrimer} holds, as claimed. 

\subsubsection{Soft constraints}\label{sec:hardsoft}

Given constraints of the form $q_i(x)=z$ (where $i$ indexes the constraints) which are imposed by a stiff potential energy, such as $U(x) = \eps^{-1}|q_i(x)-z|^2$ with $\eps \ll 1$, we can obtain the distribution for  infinitely stiff soft constraints from that for hard ones. 
This is done  by multiplying the distribution for hard constraints by a factor of $|A|^{-1/2}$, where $A$ is the Gram matrix of $\{\grad q_i\}$ evaluated at $q_i(x)=z$ \cite{Fixman:1974dd,Ciccotti:2007fv,Lelievre:2010uu}. 

If we assume the bond-distance constraints are imposed softly by spring-like forces so that $q_1(x) = |x\e{1}-x\e{2}|$, $q_2(x) = |x\e{2}-x\e{3}|$ and $z=1$, then one can calculate using \eqref{eq:param1} that  $|A|^{1/2} \propto (1+2\cos^2\omega)^{-1/2}(1+2\sin^2\omega)^{-1/2}$.  
Including this factor in \eqref{eq:pitrimer} shows the probabilities including these vibrational modes are  given by \eqref{eq:pitrimervibr}.

%\subsection{Friction by rolling? (**)}

To impose the rolling constraints softly, they must be holonomic, meaning they imply two additional constraints in configuration space only. This is the case when the rows of $C(x)$ are each a perfect gradient, but it can also hold when some nonlinear combinations of the rows are. It turns out that although each individual rolling constraint is not a perfect gradient, they are still holonomic after multiplying by a suitable integrating factor (Appendix, section \ref{sec:geometry}.) 
Therefore, with knowledge of these additional constraints one could write down the equilibrium density $\pi_{\text{roll}}$ immediately in the same way we did for $\pi_{\text{slide}}$. 
Such an approach may be able to consider larger, more general collections of discs. 

%; this is an approach we will pursue for more general collections of discs in a future publication. 

We do not attempt to impose the rolling constraints softly here, for at least two reasons. One, because it is not clear whether the additional conserved quantities in configuration space come from a stiff potential that is the origin of the friction force, or whether they are accidents of our two-dimensional geometry; this probably depends on the details of how the friction comes about. Two, because there are infinitely many functions $q_i$ which have the same level set and we currently have no physical principle with which to choose one.

\section{Discussion} \label{sec:discussion}

\subsection{Physical interpretation}

It is surprising that the equilibrium distributions for the trimer with and without rolling constraints are different, since according to classical statistical mechanics, if there is no external force on a system then each outer disc should be uniformly distributed on the central one, so the internal angle distribution is flat as for $\pi_{\text{slide}}$ in \eqref{eq:pitrimervibr}. 
What then are we to make of this result? Two interpretations are suggested here. 

First, one can take this example as a lesson in imposing constraints in a statistical mechanical system, even when these constraints are effective models for mechanical systems and the forces that impose the constraints are derived in an analogous manner to the mechanical system. Similar to the much-discussed difference between hard and soft constraints in configuration space, one must even be careful when imposing constraints on velocities, seemingly innocuous because it is not immediately obvious that these should affect distributions in configuration space. 

Nevertheless, it is often useful to model systems using constraints -- it removes fast, often unnecessary degrees of freedom, and also reduces the dimensionality, making numerical and analytical descriptions more tractable \cite[e.g.][]{HolmesCerfon:2013jw}. 
In this first interpretation where we assume the classical statistical mechanical result  holds, then this would imply a sort of ``roughness'' entropy associated with the velocity constraint, which would provide an additional force that would counteract the effect of the constraint and keep the equilibrium angle distribution constant. 
Such a roughness entropy would be similar in spirit to a vibrational entropy, but different in form because it should not necessarily be possible to obtain it as a harmonic expansion of a function of variables in configuration space only. Indeed, the constraint which models a sphere rolling on a plane is nonholonomic \cite{Johnson:2007bs,Bloch:NSbGrpH-}, so any jiggling about the constraint cannot depend only on the location and overall rotation of the sphere. Even for a pair of discs, one may wish to allow irreversible, nonharmonic slippage about their points of contact. 

To see why this suggestion is plausible, imagine the following: three gears on a slippery plane, subject to stochastic fluctuations (such as by vibrations, or fluctuations from the surrounding medium), whose centers are bound by elastic spring forces as for the trimer. The gears must roll in order to change their internal angle because the teeth are long and the spring forces strong. The teeth of the gears must have small gaps between them if the setup is to have non-zero probability, and the tangential rattling of the gears within these gaps could provide the conjectured roughness entropy in the limit as the teeth becomes smaller and closer together. 
A similar argument would hold for particles with rough surfaces, where asperities may interlock like gears with randomly-spaced teeth. The jiggling of the discs about their points of contact are coupled to the configuration space variables, since depending on the configuration (the angle of the trimer) there could be larger or smaller infinitesimal displacements available. An intriguing possibility is that this collective jiggling could  result in a roughness entropy that causes the angle distribution to deviate from a constant, or even the distribution with rolling constraints derived in this paper, since even in a classical equilibrium system it could be the case that the limiting entropy depends on the way in which the limit is obtained, i.e. whether one considers regularly spaced identical gear teeth, randomly spaced teeth with random heights, or some other pattern. The author is not aware of results showing the free energy of a collection of hard particles is a continuous function of their shape.  

Second, and perhaps more interestingly, is the literal interpretation of the result, which would imply that particles interacting with friction that creates rolling, have different free energies than those without. 
This is only possible if friction causes the system to deviate from classical statistical mechanics, which is possible if it involves non-conservative forces or kinetic effects. 
Dry friction is known to be a complicated, time-dependent, non-equilibrium phenomenon \cite{BenDavid:2010kr,Li:2011gf} that takes energy and dissipates it into heat or sound, via processes ranging from, among others, van der Waals interactions, capillary bridges, covalent bonds, plastic and elastic deformations of the bodies, fracture, wear, and quantum mechanical interactions; it is remarkable that it is so well modeled by the Coulomb interaction law across a vast range of scales \cite{Nosonovsky:2010fk,Rezek:2010ge,Vanossi:2013bt}. Yet this Coulomb interaction law involves an intrinsically nonlinear response to applied forcing and therefore is difficult to reconcile with the conditions of the fluctuation-dissipation theorem.  %\cite{Kubo:1966dq}. 
It is not so implausible that such a dissipative force could cause a system of particles to deviate from the predictions of classical statistical mechanics; indeed such deviations are observed in the widely-studied area of active particles, where active forcing due to internal motors, chemotaxis, external magnetic fields, and such forces push a system out of equilibrium  \cite[e.g.][]{Ramaswamy:2010bf,Yan:2012hg,Palacci:2013eu}. 
An active component might even be able to create a dissipative force that mimics the effects of rolling. For example, a popular method to create a reversible interaction between colloids is to coat them with strands of sticky DNA, which acts like velcro when the colloids are close enough. Certain kinds of DNA must consume fuel in order to create an effective colloid-colloid interaction, which pushes the system out of equilibrium, and could arguably cause the colloids to roll preferentially \cite{Zhang:2011hw}. 
Relatedly, colloidal particles of many different kinds are being synthesized and simulated where rotational degrees of freedom are actively forced,  
including particles that actively rotate \cite{Yan:2015ht} and look like gears \cite{Nguyen:2014dl}, for which this study may provide fundamental and preliminary intuition into a system with a rich and not very well understood phase space. 
%Finally, an exciting possibility is that robots will eventually be made on the nanoscale, where they will be subjects to mechanical constraints as well as stochastic fluctuations and may therefore be modelled in a similar way to this paper; some nanostructures have already been made that appear to roll along a surface, like driving a car \cite{Shirai:2006gd}. 

%Gears have even been used as the basis for mechanical metamaterials \cite{Meeussen:2016va}, and if these systems are made on smaller scales where thermal effects are relevant, then the calculations here will apply directly. 

If \eqref{eq:pitrimervibr} does describe the equilibrium angle distribution of a collection of particles interacting with very strong dry friction or other similar nonequilibrium dissipative forces, then it provides a method determine experimentally whether friction is present for a certain type of particle: one can construct a trimer that stays connected for long enough to generate sufficient statistics of the internal angle, and then compare the distributions. 
For example, the probability of a rolling cluster having angle greater than $\omega_c=2.2$ (where the two densities cross) is 0.48, while that for a sliding cluster is 0.45; 
measuring $P(\omega>\omega_c)$ could be one way to compare the distributions. 
Conversely, given a system where strong friction is present, measuring the angle distribution of a trimer (or other cluster of discs or spheres) could be one way to verify whether friction changes its free energy. 
It is worth noting that gears have been used as the basis for mechanical metamaterials \cite{Meeussen:2016va}, and if these systems are made on smaller scales where thermal effects are relevant, then they could be used to test (or possibly implement) the predictions in this paper, at least for certain kinds of classically-imposed velocity constraints.

\subsection{Mathematical interpretation}

Even at the mathematical level, it is perhaps surprising that the equilibrium distributions for sliding and rolling discs are different, since rolling constraints do not change the accessible configurations in position space. 
Some insight into the mathematical reason for why comes from imagining how the constraints alter the amount of white-noise forcing that is projected onto the position variables, producing observable motion. 
The white noise acts equally in all directions in the subspace spanned by the columns of $P(x)$, but the forcing we observe in the position variables depends on the projection of the noise to the subspace $\mathcal N=\{x\!\!: \theta_1\!\!=\!\!\theta_2\!\!=\!\!\theta_3\!\!=\!\!0\}$.  The magnitude of this observed forcing depends on the angles between the two subspaces, which varies with $x$. 
A stochastic process with no drift spends more time in regions where it diffuses more slowly, so the equilibrium distribution changes accordingly. 

As a side note, we can determine the specific magnitude of this projection from the calculations in section \ref{sec:derivation}. The subspace spanned by the columns of $P(x)$ has an orthonormal basis contained in the columns of $E = (t_\omega, t_\phi, t_r)\in\R^{9\times 3}$, where the vectors $t_i$ are defined in \eqref{eq:horizontal}. The subspace $\mathcal N$ where only position variables vary has an orthonormal basis contained in the columns of $F = (K^{-1}\pp{x}{\omega}, L^{-1}\pp{x}{\phi})\in\R^{9\times2}$, where $\omega,\phi$ are defined in \eqref{eq:param1} and $K(\omega),L(\omega)$ are defined in \eqref{eq:K},\eqref{eq:L}. The element of area on one subspace changes magnitude when projected to the other subspace by an amount equal to $|E^TF|=|F^TE|$ \cite{Bjorck:1973jq}, where the determinant applied to a rectangular matrix is the product of its singular values. We can calculate this determinant to be $(K^2+8)^{-1/2}(L^2+8/3)^{-1/2}$, which is consistent with the equilibrium distribution \eqref{eq:pitrimer} and also reminiscent of \eqref{eq:alph}. 

Physically, these calculations tell us how much forcing is absorbed by the spinning of the gears, and how much produces observable motion in the internal angle $2\omega$ or the overall rotation $\phi$. 
For example, consider how the cluster might change the angle $\omega$ by some small amount $\Delta \omega$. This requires a change in positions with magnitude $|\Delta \bar{x}|\approx K(\omega)|\Delta \omega|$. When discs can slide, all the white noise forcing may be applied to change the angle so the timescale for this change to happen is roughly $\Delta t \approx K^2(\omega)|\Delta \omega|^2/2$. However, if the discs must roll, then  \eqref{eq:horizontal} shows that it takes a constant amount of spinning to change the angle by some small amount $\Delta \omega$. This spinning has magnitude $|(-2,0,2)| = \sqrt{8}$ so it absorbs a constant amount of forcing, producing a timescale of $\Delta t \approx (K^2(\omega)+8)|\Delta \omega|^2/2$. The difference with the sliding case arises because to change $\omega$, the gears must spin in a way that is not proportional to how much they move in position space.

\section{Outlook and Conclusion}\label{sec:conclusion}

We have derived a set of overdamped Langevin equations for systems with linear velocity constraints. We applied this to a trimer of discs whose internal angle can change, and derived the equilibrium distribution in two cases: one where the discs can slide against each other, the other where they must roll. The two distributions are different, which shows that rolling dynamics modeled as velocity constraints can change even such basic things as the free energy of a system. 

Whether this model is physically valid depends on the details of how the friction force imposing the velocity constraint arises, a question we do not attempt to answer here since dry friction is a complicated and not fully understood phenomenon. For it to create the demonstrated free energy difference, the friction must be a nonequilibrium force, to push the system away from the classical Boltzmann equilibrium. 
Regardless of whether or not it is, this example is a useful lesson in modeling statistical mechanical systems by imposing constraints: even if the constraints act on the velocities, they can still have a fundamental effect on positions. 
We discussed how in a classical system in equilibrium, we would expect the distributions with and without rolling constraints to be the same, and conjectured that there may be a form of entropy, a ``roughness'' entropy, associated with the rolling constraints which models the infinitesimal jiggling and slippage of the discs about their points of contact as they roll around each other. Such an entropy would be similar in spirit to a vibrational entropy but structurally different, since it would be associated with the dynamical degrees of freedom and not purely with the locational ones. 

We suggested ways to test the predictions of this model via experiments on clusters of colloidal particles, or with a system of gears on a vibrating table, where
macroscale measurements like the internal configuration of a cluster may help determine microscale interactions. 
Experiments that measure the effect of friction on the steady-state properties in any system with stochastic fluctuations would be valuable, because it is clear we do not have an adequate understanding of this phenomenon which is becoming increasingly important in soft-matter and other mesoscale systems. 

Our model has also suggested problems where new mathematical developments could help shed light on physical systems.  
 Our derivation of the overdamped Langevin equations is valid for arbitrary linear velocity constraints, both holonomic and nonholonomic. In the former case the overdamped equations describe a Brownian motion on a manifold, whose equilibrium distribution is the surface measure on the manifold, but in the latter there is no such interpretation. While it turns out that discs in the plane are holonomic, a cluster of spheres should be nonholonomic: it can access a space that is higher-dimensional than the space along which it is constrained to move. This should be true because a single sphere rolling on a plane is non-holonomic \cite{Johnson:2007bs,Bloch:NSbGrpH-}. Geometrically, it lives on a sub-Riemannian manifold \cite{Montgomery:2006gl,Capogna:2007tj,Gromov:2007vn}, an object which has been little studied in the physics literature. 
In this case there is no general method to determine the equilibrium distribution of \eqref{eq:dx}, since there is no canonical volume form (surface measure) on a sub-Riemannian manifold \cite{Barilari:2013dw}. It is difficult to even identify a Laplacian, since it is not clear which volume form to use to define the divergence operator, though some recent progress has been made in comparing different choices \cite{Barilari:2013dw,Gordina:2016bc}. 
%If a suitable volume form is found, it could ... be the roughness entropy... 
One could probably work out the equilibrium distribution for individual cases directly as we have done in this paper, but the delicacy of parameterizing $SO(3)$ requires separate treatment.  Extending this study to spheres would not only potentially provide an experimental method to determine whether friction is present, but would also bring insight into the physics of stochastic, nonholonomic systems, which have rarely been considered. % outside the pure mathematics literature.%, as well as out-of-equilibrium behaviour in general. 

\bigskip

\begin{acknowledgments}
 I wish to thank Robert Kohn, Eric Vanden-Eijnden, Robert Haselhofer, Jeff Cheeger, Xue-Mei Li, Mark Tuckerman, Vinothan Manoharan, and Paul Chaikin for helpful discussions. Many thanks also to Montacer Essid for finding mistakes in previous versions of this draft. (Any remaining mistakes are purely my own.) 
This material is based upon work supported by the U.S. Department of Energy, Office of Science, Office of Advanced Scientific Computing Research under award DE-SC0012296.
\end{acknowledgments}

%%%%%%%%%%%%%%%%%%%
%%%          Bibliography           %%%
%%%%%%%%%%%%%%%%%%%

%\bibliography{BibColloids.bib,BibRolling.bib,/Users/mirandaholmes-cerfon/Dropbox/Work/Bibliographies/divergence.bib,/Users/mirandaholmes-cerfon/Dropbox/Work/Bibliographies/ColloidBib.bib}

%merlin.mbs apsrev4-1.bst 2010-07-25 4.21a (PWD, AO, DPC) hacked
%Control: key (0)
%Control: author (8) initials jnrlst
%Control: editor formatted (1) identically to author
%Control: production of article title (-1) disabled
%Control: page (0) single
%Control: year (1) truncated
%Control: production of eprint (0) enabled
%

%%%%%%%%%%%%%%%%%%%%%%
%%%       Supplementary Material        %%%
%%%%%%%%%%%%%%%%%%%%%%

\appendix

%-----------------------------------%
%\section{Appendix}

%-----------------------------------%
\section{Mass-scaled coordinates}\label{sec:mass}

We show how the mass matrix $M$ can be eliminated from \eqref{eq:lang} by a suitable change of variables. This is not a non-dimensionalization and the mass still appears implicitly in the new variables. 
Let $y = M\dot{x}$, $\tilde{x} = M^{1/2}x$, $\tilde{y} = M^{-1/2}y = M^{1/2}\dot{x}$. Then \eqref{eq:lang} becomes
\begin{align}
\dot{\tilde{x}} &= \tilde{y} \nonumber \\
\dot{\tilde{y}} + \tilde{\Gamma}\dot{\tilde{x}} &= \grad_{\tilde{x}}\tilde{U}(\tilde{x}) dt + \tilde{\sigma}\eta(t) - \tilde{C}^T\lambda. \label{eq:mass1}
\end{align}
and the constraints become
\begin{equation}
\tilde{C}(\tilde{x})\dot{\tilde{y}} = 0. \label{eq:mass2}
\end{equation}
Here 
\begin{align*}
\tilde{\Gamma}(\tilde{x}) &= M^{-1/2}\Gamma(M^{-1/2}\tilde{x})M^{-1/2}\\
\tilde{\sigma}(\tilde{x}) &= M^{-1/2}\sigma(M^{-1/2}\tilde{x})\\
\tilde{U}(\tilde{x}) &= U(M^{-1/2}\tilde{x})\\
\tilde{C}(\tilde{x}) &= C(M^{-1/2}\tilde{x})M^{-1/2}
\end{align*}
The friction and forcing remain in fluctuation-dissipation balance. Equations \eqref{eq:mass1},\eqref{eq:mass2} have exactly the same structure as \eqref{eq:lang},\eqref{eq:C} respectively, so hereafter we work in these mass-scaled coordinates and remove the tildes.

%-----------------------------------%
\section{Solving for the Lagrange multipliers}\label{sec:lagrange}

The time derivative of \eqref{eq:C} is:
\begin{equation}
C(x)\ddot{x} + \grad C(x)(\dot{x},\dot{x}) = 0,
\end{equation}
where the second term is a vector with components $(\grad C(x)(\dot{x},\dot{x}))_i = \sum_{i,j}\pp{c_i}{x_j}\dot{x}_i\dot{x}_j$. % $\grad C(x)(\dot{x},\dot{x}))_i = \dot{x}^T(\grad c_{i1},\ldots,\grad c_{im})\dot{x}$. 
Substituting for $\ddot{x}$ from \eqref{eq:lang} gives
\begin{multline}
C^T\lambda = \\
-P^\perp\Gamma\dot{x} + P^\perp \sigma \eta + P^\perp \grad U(x) + C^TG^{-1}\grad C(\dot{x},\dot{x})
\end{multline}
where $P^\perp(x)$ is the projection matrix onto the row space of $C(x)$, and $G(x)$ is the Gram matrix. Specifically:
\begin{equation}\label{eq:b3}
P^\perp = C^TG^{-1}C,  \qquad G = CC^T.
\end{equation}
One can check that $(P^\perp)^2=P^\perp$,  and $(P^\perp)^T = P^\perp$ so it is orthogonal. 
Substituting for $\lambda$ in \eqref{eq:lang} gives
\begin{multline}
\ddot{x} + P\Gamma\dot{x} =\\ -P\grad U(x) + P\sigma \eta + C^TG^{-1}\grad C(\dot{x},\dot{x})
\end{multline}
Here  $P(x) = I-P^\perp(x)$ is the projection of the velocities onto the tangent space to the manifold in phase space satisfying the constraints. One can check that $P^T=P$ so it is an orthogonal projection.

%-----------------------------------%
\section{Derivation of the overdamped dynamics}\label{sec:overdamp}

In this section we derive the equations for the dynamics in configuration space ($x$-variables only) when viscous friction is large, and over long timescales. 
Let $y = \dot{x}$,  let $\Gamma\to\Gamma/\eps$, and let $t\to t/\eps$, with $\eps\ll 1$. The equations become
\begin{align}
\dot{x} &= \frac{Py}{\eps} \nonumber\\
\dot{y} &= -\frac{\Gamma_Py}{\eps^2} + \frac{\sigma_P}{\eps}\eta - \frac{P\grad U}{\eps} + \frac{C^TG^{-1}\grad C(y, y)}{\eps} \label{eq:eps1}
\end{align}
We have defined $\Gamma_P = P\Gamma P$ and $\sigma_P = P\sigma$. 
These terms remain in fluctuation-dissipation balance, and $\Gamma_P$ is symmetric.
We can replace $y$ with $Py$, since the dynamics preserves the constraint $C(x)y=0$.

%To obtain the dynamics as $\eps\to 0$, we perform formal homogenization on the generator of the process \cite{Pavliotis:2008wg}. 
%This is a standard technique used to obtain the overdamped Langevin equations asymptotically; the only novelty here is we include arbitrary linear velocity constraints.
The backward equation for \eqref{eq:eps1} is 
\begin{equation}\label{eq:backward}
\pp{\phi}{t} = \frac{\scr{L}_0\phi}{\eps^2} + \frac{\scr{L}_1\phi}{\eps} 
\end{equation}
where 
\begin{align*}
\scr{L}_0 &= -\Gamma_Py\cdot \grad_y + \beta^{-1}\Gamma_P:\grad^2_y \\
\scr{L}_1 &= Py \cdot \grad_x - P\grad U \cdot \grad_y + C^TG^{-1}\grad C(y, y) \cdot \grad_y
\end{align*}
We write $\grad_x$, $\grad_y$ for the gradient acting only on the $x,y$ variables respectively. 

We formally expand the solution to \eqref{eq:backward} as $\phi = \phi_0 + \eps\phi_1 + \eps^2\phi_2 + \ldots$, and collect terms of the same order. The leading order equation is $\scr{L}_0 \phi_0 = 0$. 
Since $\scr{L}_0$ acts only on the $y$-variables, we must have  
\begin{equation}
\phi_0(x,y,t) = \phi_0(x,t).
\end{equation}

 The next-order equation is $-\scr{L}_0\phi_1 = \scr{L}_1\phi_0$. 
 Since $\scr{L}_0$ is linear in $y$, this is straightforward to solve, as 
 \begin{equation}
 \phi_1 = \Gamma_P^\dagger y\cdot \grad_x \phi_0,
 \end{equation}
 where $\Gamma_P^\dagger$ is the Moore-Penrose pseudoinverse of $\Gamma_P$. To check this, we calculate 
 \begin{align*}
 -\scr{L}_0\phi_1 &= \Gamma_Py\cdot \grad_y(\Gamma_P^\dagger y\cdot \grad_x\phi_0) \\
 &= \Gamma_Py\cdot \Gamma_P^\dagger \grad_x\phi_0 \\
 &= y^T\Gamma_P\Gamma_P^\dagger \grad_x\phi_0 \\
 &= y^TP\grad_x\phi_0\\
 &= Py\cdot \grad_x\phi_0 \\
 &= \scr{L}_1\phi_0
 \end{align*}
where we have used the fact that $\Gamma_P^T = \Gamma_P$, and $\Gamma_P\Gamma_P^\dagger$ is an orthogonal projection onto the column space of $\Gamma_P$ \cite{Strang:va}, so it equals $P$. %We have replaced $y$ with $Py$ in $\scr{L}_1$. 

The final equation is $-\scr{L}_0\phi_2 = -\pp{\phi_0}{t} + \scr{L}_1\phi_1$. 
By the Fredholm alternative, a solution exists only if the inner product with any element in the null space of $\mathcal{L}_0^*$ is zero. This gives the solvability condition 
\begin{equation}\label{eq:solv}
\int \pi(y)\left( -\pp{\phi_0}{t} + \scr{L}_1\phi_1\right) dy =0 ,
\end{equation}
where $\pi(y)$ is any solution to $\scr{L}_0^*\pi(y)=0$. When the integral above is explicitly evaluated, the fast variables $y$ are eliminated and we obtain an evolution equation for $\phi_0$ in the slow variables $x$. 

To calculate this integral explicitly, we first find $\pi$, which is the equilibrium distribution for the velocities (the fast variables) when the positions and spins (the slow variables) are held constant. The adjoint of $\scr{L}_0$ is
\begin{equation}
\scr{L}_0^*\rho = \beta^{-1}\grad_y\cdot \left( e^{-\half \beta|y|^2}\Gamma_P\grad_y\big( e^{\half \beta|y|^2}\rho\big)\right)
\end{equation}
We have used the fact that $\Gamma_P$ is independent of $y$, to pull it out of the inner gradient. 
It is clear that the invariant measure is 
\begin{align}
\pi(y) &= Z^{-1}e^{-\half \beta|y|^2}\sigma_{\Sigma_x}(dy) \nonumber\\
&= Z^{-1}e^{-\half \beta|y|^2}\delta(C(x)y)|G|^{1/2}, \label{eq:piy}
\end{align}
where $\sigma_{\Sigma_x}(dy)$ is the surface measure on the linear subspace $\Sigma_x(y) \equiv \{y:C(x)y=0\}$, and $Z$ is a normalization constant to ensure that $\int \pi(y)dy = 1$. 
The density must be restricted to $\Sigma_x$ since the dynamics remain on this subspace. 
We used the co-area formula 
$\sigma_{\Sigma_x}(dy)=\delta(C(x)y)|G|^{1/2}dy$  % Lelievre, p. 157
 to write \eqref{eq:piy} in both mathematicians' and physicists' notation. The matrix $G$ was defined in \eqref{eq:b3}.

\begin{widetext}
Next, we evaluate each of the terms in \eqref{eq:solv}. We have $\int \pi(y)\pp{\phi_0}{t} dy = \pp{\phi_0}{t}$. The other terms are 
\begin{equation}
\scr{L}_1\phi_1  = \underbrace{Py\cdot \grad_x(y^T\Gamma^\dagger_P\grad_x\phi_0)}_{\text{term 1}}  
 - \underbrace{P\grad_x U\cdot \Gamma_P^\dagger \grad_x\phi_0}_{\text{term 2}}
  + \underbrace{C^TG^{-1}\grad_xC(y,y)\cdot \Gamma_P^\dagger\grad_x\phi_0}_{\text{term 3}}
\end{equation}
Let's evaluate the integral of $\pi(y)$ over each of the terms in turn. We will make use of the following fact:

%LEMMA\\
\begin{equation}\label{eq:yy}
\int \pi(y)y_iy_j dy = \beta^{-1}P_{ij}.
\end{equation}

%PROOF\\
To show this, 
consider an orthonormal basis $\{e_i\}_{i=1}^d$ of the column space of $P(x)$, and let $z_i = y\cdot e_i$ be the variables lying along these directions. 
Then \begin{align*}
\int yy^Te^{-\half\beta|y|^2}\delta(C(x)y)|G|^{1/2}dy &= \int Py(Py)^Te^{-\half\beta|y|^2}\delta(C(x)y)|G|^{1/2}dy \\
&=  \sum_{k,l} e_ke_l^T \int z_kz_l e^{-\half \beta |z|^2}dz \\
&= \sum_{k,l} \beta^{-1}\delta_{kl}e_ke_l^T \\
&= \sum_{k} e_ke_k^T = P.
\end{align*}

We use \eqref{eq:yy} to calculate the integral of term 1:
\begin{equation*}
\int \pi(y)y_ky_j\partial_k((\Gamma^\dagger_P)_{ij}\partial_i\phi_0) dy = \beta^{-1}P_{kj}\partial_k((\Gamma^\dagger_P)_{ij}\partial_i\phi_0)
= \beta^{-1}\Tr(P\grad(\Gamma_P^\dagger\grad\phi_0)),
\end{equation*}
where $(\grad v)_{jk} = \partial_k v_j$. The subscript $x$ is removed on the final gradient, since it is no longer needed. 
\end{widetext}

The integral of term 2 is $\Gamma_P^\dagger P\grad_x U\cdot  \grad_x\phi_0 = \Gamma_P^\dagger \grad_x U\cdot  \grad_x\phi_0$,
since there are no terms containing $y$. This uses the fact that $\Gamma_P^\dagger P = \Gamma_P^\dagger\Gamma_P\Gamma_P^\dagger = \Gamma_P^\dagger$, by the properties of the pseudoinverse. 

Term 3 can be written as $(\grad_x\phi_0)^T\Gamma_P^\dagger C^TG^{-1}\grad C(y,y)$. But $\Gamma_P^\dagger C^T = 0$, 
since $\Gamma_P^\dagger C^T = \Gamma_P^\dagger \Gamma_P\Gamma_P^\dagger C^T = \Gamma_P^\dagger P C^T = 0$, using the properties of the pseudoinverse and the fact that the columns of $C^T$ are orthogonal to $P$. Therefore this term equals 0. 
%\begin{equation*}
%C_{lm}G_li\partial_jC_{ik}y_jy_k(\Gamma_P^\dagger)_{mn}\partial_n\phi_0
%\end{equation*}

Putting this together gives the following evolution equation for $\phi_0$:
\begin{equation}\label{eq:Lbar}
\pp{\phi_0}{t} = \Gamma_P^\dagger \grad U\cdot  \grad\phi_0 + \beta^{-1}\Tr(P\grad(\Gamma_P^\dagger\grad\phi_0)) .
\end{equation}

%-----------------------------------%
\section{Numerically simulating the Langevin equations}\label{sec:numerics}

We numerically simulated the Langevin equations \eqref{eq:lang} by writing this second-order equation as two first-order equations for the positions/spins $q = x$ and momenta $p = m\dot{x}$. We used a mass $m$ and friction coefficient $\gamma$ that were the same for all variables. 
We alternated updates of $q$, $p$ by cycling through the following four steps: 
\begin{enumerate}
\item Update $q$ by increment $\Delta q=p/m\:\Delta t$;
\item Project $q$ to manifold where bond-distance constraints \eqref{eq:con1} are exactly satisfied (the projection method was the same as that used in \cite{HolmesCerfon:2013jw});
\item Update $p$ by increment $\Delta p= -(\gamma/m)p\Delta t + \sigma \sqrt{\Delta t}\:N$, where $N\in \R^{9}$ is a vector of independent standard normal random variables;
\item Project $p+\Delta p$ to space of allowed velocities (this is done by multiplying by matrix $P$ defined in \eqref{eq:P}.)
\end{enumerate}
Apart from the projection step, this is exactly an Euler-Maruyama method so is expected to be weakly first-order accurate \cite{Kloeden:1992hi}. 
We did not include a non-overlap condition for the discs, though this is easily accounted for a-posteriori by truncating the histogram. 
The parameters used were $m=0.1$, $\gamma = 1$, $\sigma=1$. We set $\Delta t = 5\times10^{-3}$ for the sliding simulations, and $\Delta t = 1\times10^{-4}$ for the rolling ones. 
A finer timestep was needed for the rolling simulations to get good agreement with the theory, presumably because the simulations do not conserve the additional implied conserved quantities in configuration space \eqref{eq:Qtrimer} (see section \ref{sec:geometry}.) 
The total time each simulation was run for was $T_{max}=10^5$ for sliding discs and  $T_{max}=1.8\times10^5$ for rolling discs. 
We needed to run the rolling simulations longer than the sliding ones to converge to the equilibrium distribution, because the effective diffusion coefficient in angle space is smaller.

%-----------------------------------%
\section{Geometry of the rolling trimer's configuration space}\label{sec:geometry}

To calculate the trimer's equilibrium distribution, we did not need to know the geometric structure of its configuration space-- neither the constants of integration nor the dimension of the manifold on which it lives. The calculation was possible because of the symmetries that let us project the dynamics to a lower-dimensional manifold without losing information, and on this lower-dimensional manifold the trimer had no constraints.
Nevertheless, this geometric structure is an interesting mechanics problem in itself. 

Let us count degrees of freedom: we began with five parameters to describe the configuration space, and two constraints, so there is a three-dimensional space along which the cluster can move (the ``horizontal space.'') What is the actual dimension of the space in which it lives? 

This can be understood  by calculating iterated Lie brackets of the horizontal space. It is simplest to do this in the parameterized space, in which an orthogonal basis of horizontal tangent vectors (proportional to $t_\omega, t_\phi, t_r$) is
\begin{align}
T_\omega &= (1,0,-2,0,2),\nonumber\\
T_\phi &= (0,3,2,4,2), \nonumber \\
 T_r &= (0,0,1,-1,1).  \label{eq:T}
\end{align}
(See section \ref{sec:tangentmap} for an explanation.)
In this parameterization the horizontal space is a single, constant plane; clearly all Lie brackets give 0. Therefore by the Frobenius theorem \cite{lee} the trimer lives on a three-dimensional manifold, so there are two extra conserved quantities. 
%We can find these by finding a basis for the normal space. 
%One way to do this is to choose vectors whose $\theta$-components equal those of $T_\omega$, $T_\phi$, and choose the corresponding non-zero spatial component to make them perpendicular to those vectors. 
%This gives the following two normal vectors (after dividing by common factors):
One can check that a basis for the normal space is 
\begin{align}
N_1&= (-4,0,-1,0,1),\nonumber\\
N_2 &= (0,-4,1,2,1).
\end{align}
These are the gradients of the following scalar functions: 
\begin{align}
Q_1 &= -4\omega - \theta_1 + \theta_3,\nonumber\\
Q_2 &= -4\phi + \theta_1 + 2\theta_2 + \theta_3. \label{eq:Qtrimer}
\end{align}
It is these functions (or any nonlinear function of them) which are conserved by the dynamics with rolling constraints. 
From these one could calculate the equilibrium distribution directly. 
%Note however that it is not always straightforward to find the conserved quantities, since, as was the case here, these do not come from directly integrating the constraints. 
%We plan to discuss the geometry of rolling discs in more depth in a future publication. 

\subsubsection{Tangent map}\label{sec:tangentmap}

The horizontal vectors \eqref{eq:horizontal} 
come from considering the \emph{tangent map} induced by a smooth map $f:M\to N$ from one manifold $M$ to another manifold $N$. Recall that a tangent vector at point $p$ can be thought of as an equivalence class of curves $[c]$, where the equivalence relation is $c(t) \equiv d(t)$ if $c'(0) = d'(0)$ and $c(0) = d(0) = p$ \cite{Kuhnel:2002vh}. Then, $f$ induces a natural linear map between tangent spaces, $df_p:T_pM\to T_pN$, defined by 
\[
df_p(c'(0)) = \dd{}{t}\Big|_{t=0}f(c(t)).
\]
If we have a description of the manifolds in the variables $x\in X$, $y\in Y$ where $X,Y$ are subsets of suitable spaces, and if we have a map $f:X\to Y$, then the tangent map is 
\begin{align*}
df_p(c'(0)) &= \grad f \: c'(0) \\
&= \begin{pmatrix}
\pp{x_1}{y_1} &  \pp{x_1}{y_2}\ & \cdots &\pp{x_1}{y_n} \\
\pp{x_2}{y_1}& \pp{x_2}{y_2} & \cdots &\pp{x_2}{y_n} \\
\vdots &&& \vdots
\end{pmatrix}
\begin{pmatrix}c_1'(0)\\c_2'(0) \\ \cdots \\ c_n'(0)\end{pmatrix}
\end{align*}

Let $M$ be the manifold of accessible configurations parameterized by the variables $(\omega,\phi,\theta)$, and let $N$ be the same manifold described by the variables $x$. 
We have an explicit mapping $f:M\to N$, given by \eqref{eq:param1} and the subsequent inline equations. The Jacobian of this mapping in block form is 
\begin{equation}
\grad f = \begin{pmatrix}
\pp{\bar{x}}{\omega} & \pp{\bar{x}}{\phi} & 0 \\
0 & 0 & I_3
\end{pmatrix},
\end{equation}
where $I_3$ is the $3\times3$ identity matrix and $0$ is a matrix of zeros with dimensions correct for the context.  
From this, we can see that 
\[
\grad f\: T_\omega \;\propto\; t_\omega, \quad
\grad f\: T_\phi \;\propto\; t_\phi,\quad
\grad f \;T_r \;\propto\; t_r.
\]

\end{document}